\documentclass[preprint,flushrt]{aastex}
\slugcomment{ApJ, accepted}
\usepackage{lineno}
\linenumbers

\def\ltsima{$\; \buildrel < \over \sim \;$}
\def\simlt{\lower.5ex\hbox{\ltsima}} 
\def\gtsima{$\; \buildrel > \over \sim \;$}
\def\simgt{\lower.5ex\hbox{\gtsima}} 

\def\deg{\hbox{$^\circ$}}
\def\gunit{$\times$ 10$^{-7}$ ph cm$^{-2}$ s$^{-1}$}
\def\r95{$r_{\rm 95}$}
\def\swift{{\em Swift}}

\newcommand{\G}{\Gamma}
\newcommand{\g}{\gamma}
\newcommand{\cm}{\mathrm{cm}}

\newcommand{\erg}{\mathrm{erg}}
\newcommand{\eV}{\mathrm{eV}}
\newcommand{\keV}{\mathrm{keV}}
\newcommand{\MeV}{\mathrm{MeV}}

\newcommand{\s}{\mathrm{s}}
\newcommand{\dday}{\mathrm{day}}

\newcommand{\Mpc}{\mathrm{Mpc}}

\shorttitle{{\em Fermi}-LAT Observations of Cen A}
\shortauthors{Abdo et al.}

\begin{document}

\title{{\em Fermi} Large Area Telescope View of the Core of the Radio 
Galaxy Centaurus~A}
\author{
A.~A.~Abdo\altaffilmark{1,2}, 
M.~Ackermann\altaffilmark{3}, 
M.~Ajello\altaffilmark{3}, 
W.~B.~Atwood\altaffilmark{4}, 
L.~Baldini\altaffilmark{5}, 
J.~Ballet\altaffilmark{6}, 
G.~Barbiellini\altaffilmark{7,8}, 
D.~Bastieri\altaffilmark{9,10}, 
B.~M.~Baughman\altaffilmark{11}, 
K.~Bechtol\altaffilmark{3}
% ...
}
\altaffiltext{1}{Space Science Division, Naval Research Laboratory, Washington, DC 20375, USA}
\altaffiltext{2}{National Research Council Research Associate, National Academy of Sciences, Washington, DC 20001, USA}
\altaffiltext{3}{W. W. Hansen Experimental Physics Laboratory, Kavli Institute for Particle Astrophysics and Cosmology, Department of Physics and SLAC National Accelerator Laboratory, Stanford University, Stanford, CA 94305, USA}
\altaffiltext{4}{Santa Cruz Institute for Particle Physics, Department of Physics and Department of Astronomy and Astrophysics, University of California at Santa Cruz, Santa Cruz, CA 95064, USA}
\altaffiltext{5}{Istituto Nazionale di Fisica Nucleare, Sezione di Pisa, I-56127 Pisa, Italy}
\altaffiltext{6}{Laboratoire AIM, CEA-IRFU/CNRS/Universit\'e Paris Diderot, Service d'Astrophysique, CEA Saclay, 91191 Gif sur Yvette, France}
\altaffiltext{7}{Istituto Nazionale di Fisica Nucleare, Sezione di Trieste, I-34127 Trieste, Italy}
\altaffiltext{8}{Dipartimento di Fisica, Universit\`a di Trieste, I-34127 Trieste, Italy}
\altaffiltext{9}{Istituto Nazionale di Fisica Nucleare, Sezione di Padova, I-35131 Padova, Italy}
\altaffiltext{10}{Dipartimento di Fisica ``G. Galilei", Universit\`a di Padova, I-35131 Padova, Italy}
\altaffiltext{11}{Department of Physics, Center for Cosmology and Astro-Particle Physics, The Ohio State University, Columbus, OH 43210, USA}
% ...

\begin{abstract}

We present $\g$-ray observations with the Large Area Telescope on
board the {\em Fermi} Gamma-Ray Telescope of the nearby radio galaxy
Centaurus~A.  The previous EGRET detection is confirmed, and the
localization is improved using data from the first 10 months of {\em
Fermi} science operation.  In previous work, we presented the
detection of the lobes by the LAT; in this work, we concentrate on the
$\g$-ray core of Cen~A.  Flux levels as seen by the LAT are not
significantly different from that found by EGRET, nor is the extremely
soft LAT spectrum ($\G=2.67\pm0.10_{stat}\pm0.08_{sys}$ where the
photon flux is $\Phi\propto E^{-\G}$).  The LAT core spectrum,
extrapolated to higher energies, is marginally consistent with
the non-simultaneous HESS spectrum of the source.  The LAT
observations are complemented by simultaneous observations from {\em
Suzaku}, the {\em Swift} Burst Alert Telescope and X-ray Telescope,
and radio observations with the Tracking Active Galactic Nuclei with
Austral Milliarcsecond Interferometry (TANAMI) program, along with a
variety of non-simultaneous archival data from a variety of
instruments and wavelengths to produce a spectral energy distribution
(SED).  We fit this broadband data set with a single-zone
synchrotron/synchrotron self-Compton model, which describes the radio
through GeV emission well, but fails to account for the
non-simultaneous higher energy TeV emission observed by HESS from
2004-2008.  The fit requires a low Doppler factor, in contrast to BL
Lacs which generally require larger values to fit their broadband
SEDs.  This indicates the $\g$-ray emission originates from a slower
region than that from BL Lacs, consistent with previous modeling
results from Cen~A.  This slower region could be a slower moving layer
around a fast spine, or a slower region farther out from the black
hole in a decelerating flow.  The fit parameters are also consistent
with Cen A being able to accelerate ultra-high energy cosmic-rays, as
hinted at by results from the Auger observatory.

\end{abstract}


\section{Introduction}
\label{section-intro}

Radio galaxies exhibiting jets which terminate in radio lobes on tens
of kpc to Mpc scales are classified based on their radio morphology
and power by \citet{fanaroff74}.  They are divided into Fanaroff-Riley
(FR) type I and type II, where type I sources have the highest surface
brightness feature at the center, while in type II sources it is
farther from the core.  Furthermore, the transition radio
luminosity between FRI and FRII increases with the optical luminosity
of the host galaxy \citep{ledlow96}.  In the AGN unification scheme, blazars
are thought to be radio galaxies with the jet aligned along our
line of sight, and are subdivided into flat spectrum radio quasars
(FSRQs) and BL Lacertae objects based on the strength of emission
lines in their spectrum, where FSRQs generally have strong emission
lines, while BL Lacs have weak or none
\citep[][]{stritt72,marcha96,landt04}.  FRI galaxies are thought to
correspond to misaligned BL Lacs, while FRIIs correspond to misaligned
FSRQs \citep[e.g.,][and references therein]{urry95}, although
there is evidence that this unification scheme is too simple
\citep[e.g.,][]{landt08}.  Apparent superluminal motion observed on
milli-arcsecond size scales indicates that their jets must be moving
at high relativistic speeds, with bulk Lorentz factor $\G_j\sim$
10--20 for FSRQs and BL Lacs \citep{kellerman04,lister09}, although
some TeV BL Lacs have $\G_j\sim3$ \citep{piner08}.  The existence of
high energy and very high energy (VHE) $\g$-rays observed from these
sources provides further evidence for highly relativistic flows, as
they are necessary to avoid $\g$-ray attenuation by electron-positron
pair production \citep{dondi95}.  Indeed, this sometimes gives values
of $\G_j$ greater than that found from very-long baseline
interferometry (VLBI) superluminal observations; e.g., $\Gamma_j \ga
50$ is required for a recent outburst from PKS 2155-304
\citep[e.g.,][]{begelman08,finke08}.

Since blazars are strong sources of beamed $\g$-rays, it is natural to
think that radio galaxies may be also.  Several radio galaxies were
detected by EGRET: \object{3C~111} \citep{har08}, \object{NGC~6251}
\citep{muk02}, and Centaurus (Cen) A \citep{sre99,har99}.  The
identifications were rather uncertain, due to the large EGRET error
circles.  Only two radio galaxies have been detected so far with the
latest generation of TeV atmospheric Cherenkov telescopes,
\object{M87} \citep{aharonian06,acciari08,albert08,acciari09_m87} and
Cen~A \citep{aha09}.  The Radio Galaxy 3C~66B seems to have been seen
by MAGIC \citet{aliu09}, although the detection is questionable due to
its proximity to the BL Lac 3C~66A and its lack of detection by
VERITAS \citep{acciari09_3c66a}.  The {\em Fermi}-LAT collaboration
has reported the detections of \object{NGC~1275}
\citep[Per~A;][]{pera}, M87 \citep{m87}, and Cen~A \citep{lbas}.
Several more $\g$-ray detections of radio galaxies have been reported
in the first {\em Fermi}-LAT catalog \citep[1FGL;][]{elevenmonth,1lac}
and a future publication will examine them in more detail (Fermi
Collaboration 2010, in preparation).

The {\em Fermi} Gamma Ray Space Telescope was launched on 2008 June 11
and contains the Large Area Telescope (LAT), a pair conversion
telescope which has a field of view of about 20\% of the sky at
20~MeV to over 300~GeV \citep{atwood09}.  For the first year of
operation, {\em Fermi} was operated in a sky-survey observing mode,
wherein the LAT sees every point on the sky every $\sim3$ hours.

During the first 3-months of science operation, the {\em Fermi}-LAT
confirmed \citep{bsl,lbas} the EGRET detection of Cen~A.  Here with
additional monitoring, we present accumulated data after 10 months of
operation.  The new LAT observations bridge the gap between EGRET and
HESS, providing a detailed look at the $\gamma$-ray spectrum essential
for addressing emission models.  In addition to the LAT $\g$-ray
source in the central few kpc (hereafter the $\g$-ray ``core''),
$\g$-rays from the giant lobes of \object{Cen~A} have also been seen
with {\em Fermi}, with the origin likely to be Compton scattering of
the cosmic microwave background (CMB) and extragalactic background
light (EBL), confirming the predictions of \citet{cheung07} and
\citet{hardcastle09}.  Detailed work on separating the core and lobe
emission is presented elsewhere \citep[][hereafter referred to as the
lobe paper]{cena_lobe}, although we provide a summary of LAT
observations below. For the purposes of this paper, which is a study
of $\g$-ray emission of the core, the lobes are essentially background
sources.

We present a summary of Cen~A and observations of this object in \S\
\ref{cena-section}.  The observations of the core of Cen~A with the
LAT over the first 10 months of {\em Fermi} operation are presented in
\S\ \ref{latobs}.  We also present simultaneous Cen A core
observations from {\em Suzaku} and {\em Swift}, and radio data from
the TANAMI program in \S\ \ref{obs}.  In \S\ \ref{model-section} we
combine these with archival data and model its SED of the Cen~A core.
In \S\ \ref{discussion} we discuss the implications in detail, and we
conclude with a brief summary (\S\ \ref{summary}).

\section{Centaurus A}
\label{cena-section}

The FRI \object{Cen~A} is the nearest radio lound active galaxy
to Earth, making it an excellent source for studying the physics of
relativistic outflows and radio lobes.  Indeed, it is near enough
that its peculiar velocity dominates over the Hubble flow, and its
redshift ($z=0.00183$) cannot be used to accurately calculate its
distance.  \citet{fer07} have found that the average of several
distance indicators gives $D=3.7\ \Mpc$, which we adopt.  At this
distance, an arcsecond corresponds to about 18 pc.  Due to its
proximity to Earth, it has been well studied throughout the
electromagnetic spectrum, from radio to $\g$-rays.  Recently, the
Auger collaboration reported that the arrival directions of the
highest energy cosmic rays ($\ga 6\times10^{19}$ eV) observed by the
Auger observatory are correlated with nearby AGN, including Cen~A
\citep{abraham07,abraham08}, while \citet{moskalenko09} found that,
if the giant lobes are taken into account, as many as four ultra-high
energy cosmic rays (UHECRs) may be associated with this source.  
Although the overall significance of this correlation is reduced in
the expanded Auger data set, the significance remains high in the
direction of Cen~A \citep{abraham09}.  This suggests that
\object{Cen~A}---and other radio galaxies---may be sources of UHECRs.

\object{Cen~A} has interesting radio structure on several size scales.
The most prominent features are its giant radio lobes, which subtend
$\sim10$\deg\ on the sky, oriented primarily in the North-South
direction.  They have been imaged at 4.8 GHz by the Parkes telescope
\citep{junkes93} and studied at up to $\sim$60 GHz by
\citet{hardcastle09} utilizing Wilkinson Microwave Anisotropy Probe
\citep[WMAP;][]{hinshaw09} observations.  The North lobe contains a
bright region a few tens of arcminutes in size often referred to as
the Northern middle lobe \citep{morganti99}.  Mis-aligned by
approximately $45\deg$ relative to the outer lobes are inner radio
lobes on an arcminute scale \citep{burns83}.  A strong,
well-collimated jet can be seen on the arcsecond size scale in the
radio, and {\em Chandra} can resolve X-ray emission from it, which is
likely caused by synchrotron emission \citep{kraft02,hardcastle03}.
The innermost region of Cen~A has been resolved with VLBI, and shown
to have a size of $\sim3\times10^{16}$ cm
\citep{kellerman97,horiuchi06}.  Observations at shorter wavelengths
also reveal a small core, namely VLT infrared interferometry which
resolves the core size to $\sim 6\times10^{17}$ cm \citep{mei07}.
VLBI images reveal a weak counter jet on the milli-arcsecond scale
\citep{jones96}.  Based on the motion of the VLBI blobs, and assuming
the brightness differences of the different jets are due to Doppler
effects, \citet{tingay98} estimate the angle of the sub-parsec jet to
our line of sight to be $\sim 50-80\deg$.  Applying a similar
technique to the 100 pc scale jet which has a larger jet-counterjet
ratio, \citet{hardcastle03} estimate a jet angle of $\sim 15\deg$.
\citet{hardcastle03} speculate that the conflicting angle estimates
may be due to the assumption that the jet--counter jet brightness
differences are caused by Doppler beaming rather than intrinsic
differences.

\object{NGC~5128}, the giant elliptical host galaxy of \object{Cen~A},
contains a kiloparsec-scale dust lane.  This feature appears to be an
edge-on disk obscuring the central region and nucleus, and is probably
the remnant of a previous merger \citep{quillen92,isr98}.  It also has
a dusty torus within 100 pc of the black hole, with a high column
density ($N_H\ga10^{22}\ \cm^{-2}$ ) \citep{isr08,wei08}.  X-ray
spectra taken at various times over decade timescales indicate a
time-varying absorbing column density, which could be due to
variations in a warped disk viewed edge-on \citep{rot06}.  Estimates
for the mass of the supermassive black hole at the center of
\object{Cen~A} range from $(0.5-1) \times 10^8 M_\odot$
\citep{silge05,mar06,neu07} based on the kinematics of stars, as well
as H$_2$ and ionized gas.

With the $Compton$ Gamma-Ray Observatory, emission was detected by
OSSE \citep{kinzer95} and COMPTEL \citep{ste98} at 100s of keV to MeV
energies.  \citet{kinzer95} suggested the hard X-ray emission from
\object{Cen~A} detected with OSSE was the result of Compton-scattered
disk radiation by a thermal plasma (i.e., a hot corona), due to a
turnover in the spectrum at a few hundred keV.  However, \citet{ste98}
noted that the high-energy portion of the OSSE spectra smoothly
connected with the higher energy COMPTEL spectra, and the OSSE and
COMPTEL variability seem to be correlated.  They used this to argue
for a nonthermal jet origin for the X-rays.  \citet{eva04} have
resolved the arcsecond-scale core of Cen~A with {\em Chandra} and {\em
XMM-Newton}.  The 2--7 keV X-ray continuum, when corrected for
absorption, is consistent with what is predicted from a correlation
between unresolved X-ray emission and 5 GHz core emission for jets of
radio galaxies \citep{canosa99}.  They thus consider it likely that
nonthermal emission from the sub-pc (sub-mas) scale jet is the origin
of the continuum X-rays from the core of Cen~A.  However, hard X-rays
observed by {\em Suzaku} do not seem to fit on the \citet{canosa99}
correlation, possibly indicating a non-jet origin \citep{mar07}.  The
nature of the continuum X-ray emission from the core of Cen~A remains
an open question.

Cen~A has been a target of $\gamma$-ray observations dating back to
the 1970s \citep[e.g.,][]{gri75,hal76}.  Cen~A was seen by EGRET up to
GeV energies \citep{sre99,har99}.  The $\g$-rays are thought to
originate from a relativistic jet near the central elliptical galaxy
(the radio ``core'') analogous to blazars, although it has been
suggested that Compton-scattering of the CMB and the infrared-optical
EBL in the giant radio lobes could be a source of $\g$-rays from
\object{Cen~A} \citep{hardcastle09,cheung07} and other radio galaxies
such as \object{Fornax~A} \citep{georgan08}.  At the highest, TeV
energies, a detection was recently reported from Cen~A by the
air Cherenkov detector HESS \citep{aha09}.



\section{ {\em Fermi}-LAT Gamma-Ray Observations}
\label{latobs}

\subsection{Localization}

The EGRET detection of Cen~A \citep{sre99,har99} was confirmed early
on by the {\em Fermi}-LAT.  Based on 3-months of all-sky survey data,
the initial LAT detection was reported in the LAT bright source list
(BSL) paper \citep{bsl} as 0FGL J1325.4--4303 with a 95$\%$ confidence
localization, \r95=0.304\deg = 18.3\arcmin.  In the companion LAT
Bright AGN Sample paper \citep[LBAS;][]{lbas} to the BSL, a single
power-law fit was reported, which gave $F$($>$100 MeV) = 2.15 ($\pm$
0.45) \gunit\ with photon index, $\Gamma$ = 2.91 $\pm$ 0.18, and a
peak flux on a $\sim$1 week timescale of (3.23 $\pm$ 0.80) \gunit.
Note that this only considered the $\g$-ray emission from Cen~A as a
single point source, i.e., it did not account for any lobe emission.

\begin{figure}
\epsscale{0.7} 
\plotone{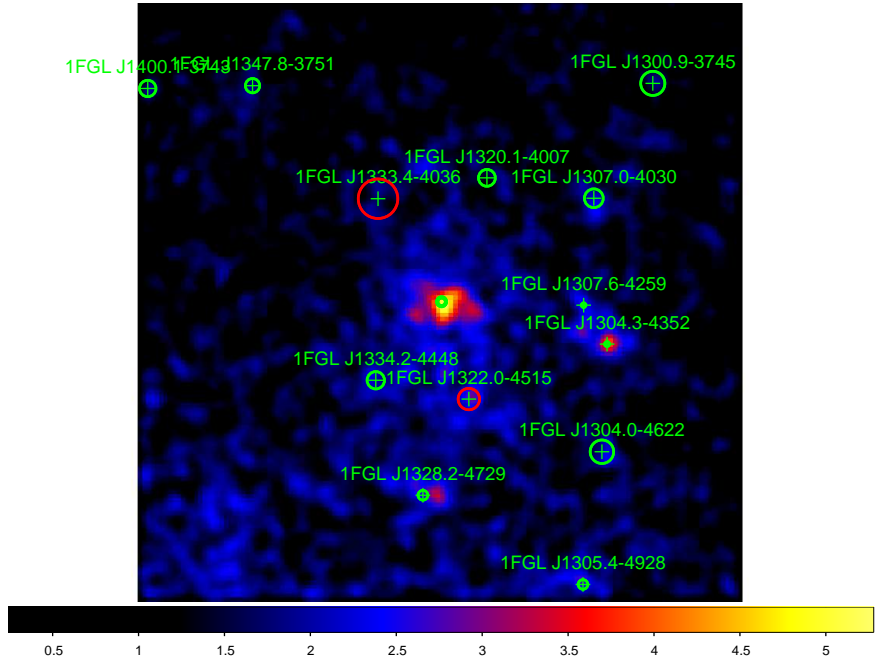} 
\caption{LAT gamma-ray image in the 0.2--30 GeV range in a
$14\deg\times14\deg$ region, smoothed by a Gaussian with
$\sigma=0.3\deg$.  The green crosses are the source in the 11 month
LAT source list.  Green circles are sources considered in the likelihood
fitting for model B (see the lobe paper).  Red circles are
additional sources considered in model A.  Circle radii represent the
semi-major error radius in the 11-month catalog. }

\label{figure-roi}
\end{figure}

To these initial observations, 7 additional months of all-sky survey
data are added to the current analysis. Specifically, the observations
span the time period from 2008 August\ 4 to 2009 May\ 31,
corresponding to MET (mission elapsed time) 239557420 -- 265507200.
Diffuse event class ({\tt CTBCLASSLEVEL}=3) events were selected with
a zenith angle cut of $<$105\deg, and a rocking angle cut of 39\deg.
The former are well calibrated and have minimal background while the
latter greatly reduce Earth albedo $\gamma$-rays. For the analysis,
LAT Science
Tools\footnote{\url{http://fermi.gsfc.nasa.gov/ssc/data/analysis/scitools/overview.html}}
version v9r11 was utilized with the {\tt P6\_V3\_DIFFUSE} instrument
response function (IRF).  The standard LAT Galactic emission
model, {\tt
GLL\_IEM\_V02.FIT}\footnote{\url{http://fermi.gsfc.nasa.gov/ssc/data/access/lat/BackgroundModels.html}}
was used and the uniform background was represented by the isotropic
diffuse $\g$-ray background and the instrumental residual background
\citep[{\tt isotropic\_iem\_v02.txt}, ][]{background}.  We consider 11
point sources in the 1FGL catalog \citep[][see also
Figure~\ref{figure-roi}]{elevenmonth}.

\begin{figure}
\epsscale{0.7} 
\plotone{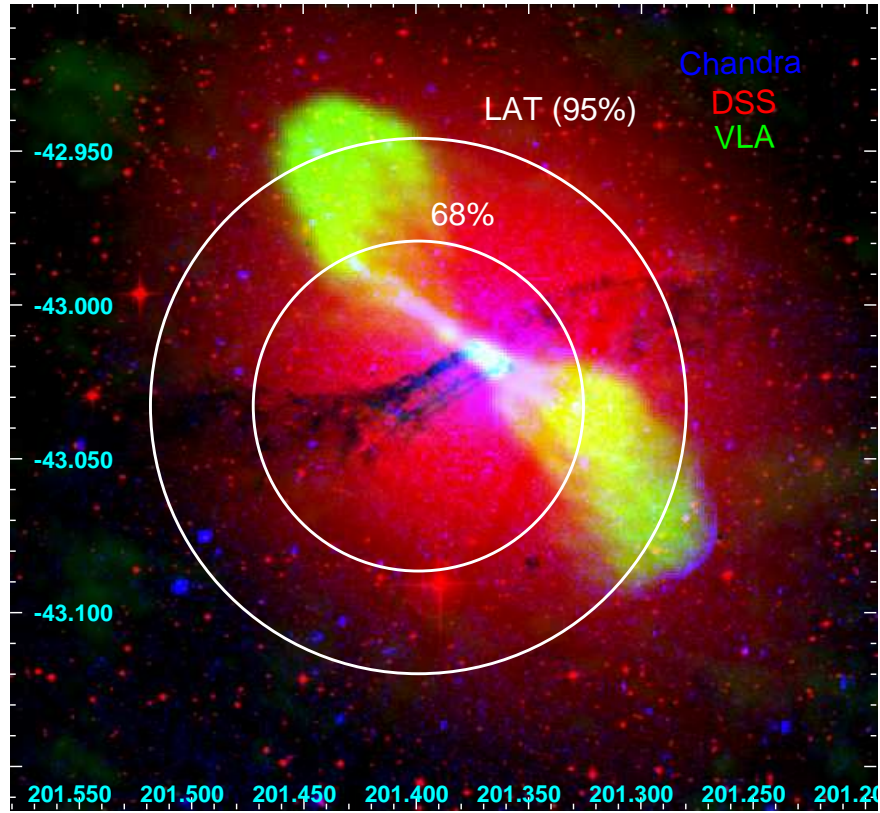} 
\caption{The LAT localization error circles indicated on a 3-color
image of Cen~A. The image is made with the VLA 21 cm image from
\citet{con96}, the optical from Digital Sky Survey plates from the
UK 48-inch Schmidt telescope, and an archival Chandra X-ray exposure
from \citep[][OBSID 7797]{hardcastle07}. The $\gamma$-ray source is
clearly positionally coincident with Cen~A, enclosing the core,
kpc-scale jet, and most of the radio lobes.  }
\label{rgb}
\end{figure}

Figure~\ref{figure-roi} shows the the 0.2$-$30 GeV LAT image
centered on Cen~A, which is clearly detected.  Also prominent is the
Galactic emission toward the south, and several faint sources in the
field.  We obtained a localization of the source at Cen A with 
{\tt gtfindsrc}, which finds point source locations based on an
unbinned likelihood analysis.  The resulting localization was
reduced to \r95= 0.087\deg\ = 5.2\arcmin\ (5.7 kpc), centered at RA =
201.399\deg, Dec = $-43.033$\deg\ (J2000.0 epoch) which is offset by
0.029\deg\ = 1.7\arcmin\ (1.9 kpc) from the VLBI radio position of
\object{Cen~A} \citep{ma98}.  Figure~\ref{rgb} shows the localization
error circle of the LAT emission overlaid on the combined radio,
optical, X-ray images.  The new LAT position is consistent with that of
3EG J1324-4314 \citep{sre99,har99}, but both are notably offset from
EGR~J1328-4337, the closest EGRET source in the \citet{cas08}
catalog. The latter derived position shifted in such a way that Cen~A
was outside of the \r95\ localization circle, so that there was
some ambiguity as to whether EGRET was actually detecting Cen~A, but
the new LAT position confirms the earlier 3EG result.  The LAT
significantly improves upon the previous EGRET $\gamma$-ray
localization (\r95=0.53\deg = 32\arcmin).


\subsection{Spatial and Spectral Analysis}
\label{LATspectralsection}

The binned likelihood fitting was performed with the {\tt gtlike}
tool, first assuming Cen~A is a point source, i.e., that there is no
$\g$-ray lobe emission (model A).  The field point source positions
were fixed, and their spectra were assumed to be power-laws, with the
photon indices allowed to vary.  The location of Cen~A was fixed at
its VLBI radio position \citep{ma98}.  In addition to the 11 1FGL
point sources used in the lobe paper, in order to treat the lobe
emission as a background source, we include two 1FGL sources,
1FGL~J$1322.0-4515$ and 1FGL~J$1333.4-4036$, which are thought to be
the local maxima of the lobe emission.  A likelihood analysis with the
energy information binned logarithmically in 20 bins in the 0.2--30
GeV band, and the $\g$-ray directions binned into a
$14\deg\times14\deg$ grid with a bin size of $0.1\deg \times0.1\deg$.
For both the Galactic and isotropic emission models, one free
parameter was introduced to adjust the normalization.  Because
the effective area of the LAT is rapidly changing below $\sim 200\
\MeV$, we use events with energy above this value.  Above 30 GeV the
significance of detection is $<3\sigma$, so we make a cut as this
energy as well.

As a result, the test statistic \citep[TS;][]{mat96} is found to
be 378 for Cen~A, which is smaller than the TS=628 in the 1FGL catalog
\citep{elevenmonth}, since the lower energy limit is 200 MeV in our
analysis, instead of 100 MeV in the catalog.  The relative
normalizations of the Galactic and isotropic models become 
1.02$\pm$0.02 and 1.40$\pm$0.06, respectively, and the fit is
reasonable within the current background model uncertainty.  This fit
gives a power-law photon index of Cen~A between 200 MeV and 30 GeV of
$\Gamma$=$2.76\pm0.07$ and the flux extrapolated down to $>$100 MeV is
(2.06$\pm$0.20) \gunit (where errors are statistical only).  As noted
in \citet{lbas}, the spectrum is very steep in comparison to the
typical blazars of $\Gamma=1.5-2.5$ The power-law photon index is
consistent with the 3EG result of $\Gamma = 2.58\pm0.26$
\citep{har99}.  The 3EG flux was reported to be ($1.36 \pm0.25$)
\gunit, and have a peak value of ($3.94\pm1.45$) \gunit\
\citep{har99}, consistent with with the average flux.  

\begin{figure}
\epsscale{1.0}
\plotone{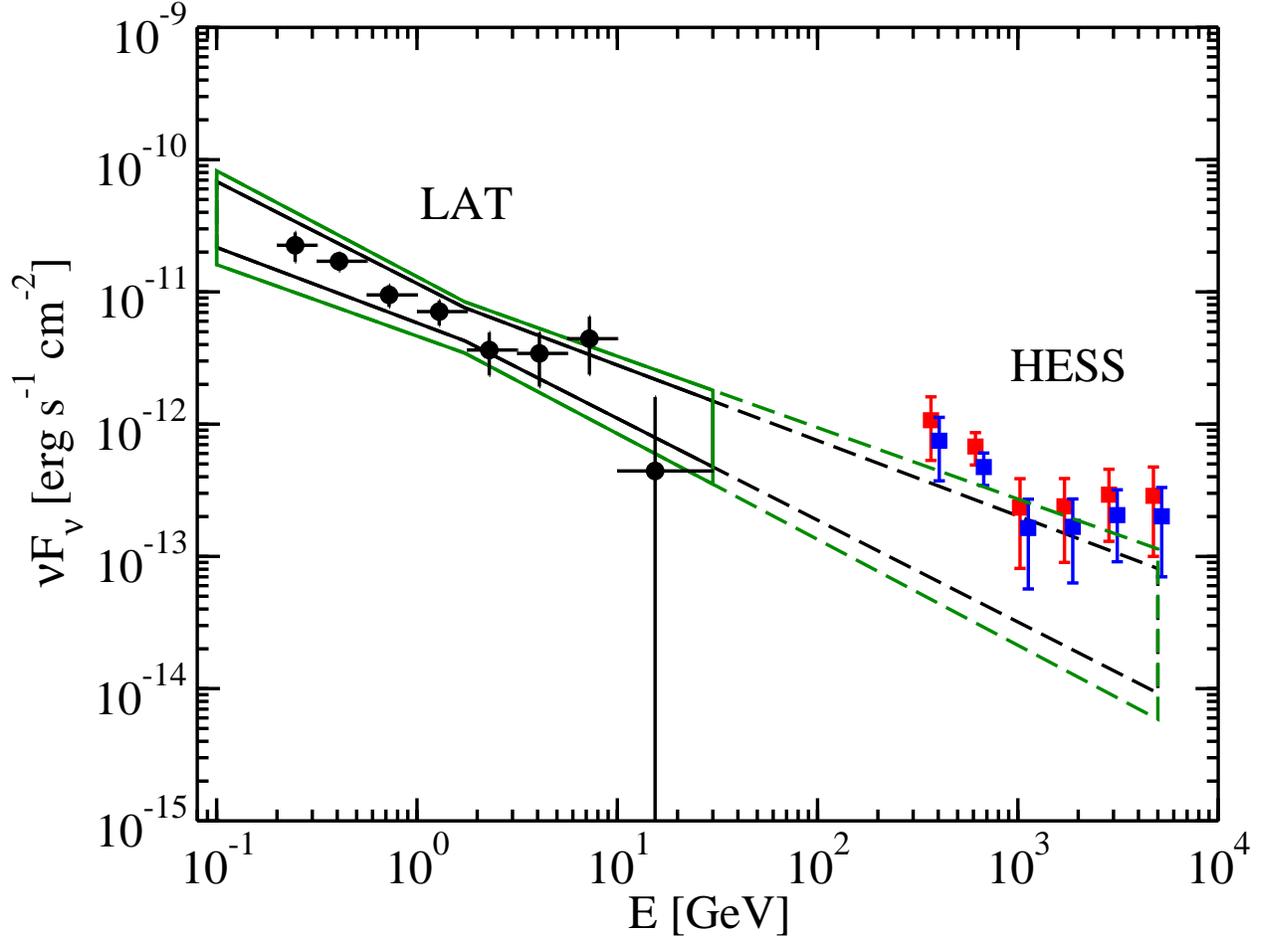}
\caption{ Spectrum of the Cen A core from differential fluxes derived
for successive energy ranges from model B (black circles).  The black
bowtie indicates the best fit 0.1 -- 30 GeV LAT flux and $\G$ with
statistical errors only, while the green bowtie indicates this with
systematic errors as well.  The LAT spectrum is extrapolated into the
HESS energy range (dashed lines).  The HESS data from \cite{aha09} are
shown (red squares) and the HESS data shifted to lower flux by their
statistical and systematic normalization error (blue squares).  The
latter are also shifted in energy by 10\% for clarity.  }
\label{graySED}
\end{figure}

We next modeled the region with a radio image of the giant lobe 
(model B).  This analysis is identical to that described in the
lobe paper, and the reader is referred to it for details.  We present
a brief description below.  We use the WMAP image at 20 GHz from
\citet{hardcastle09}, and eliminate the Cen~A core region with a cut
radius of $1\deg$.  In this analysis, we exclude two point sources
(1FGL~J$1322.0-4515$ and 1FGL~J$1333.4-4036$), which are assumed to be
emission from the lobes.  The binned likelihood analysis was performed
to extract the flux and spectral indices for the core and lobes.  The
relative normalizations of the Galactic and isotropic models become
1.00$\pm$0.02 and 1.44$\pm$0.06, respectively.  The $\g$-ray detection
in each energy range is significant at a 4$\sigma$ level up to the
5.6--10 GeV energy bin for the core region and the spectrum is
consistent with the power-law model.  This fit gives a photon index of
the core between 200 MeV and 30 GeV of
$\Gamma$=$2.67\pm0.10_{stat}\pm0.08_{sys}$ and a flux extrapolated
down to $>$100 MeV of (1.50$\pm0.25_{stat}\pm0.37_{sys}$) \gunit, with
statistical and systematic errors reported.  Here, we consider the
systematic errors from the effective area, the diffuse model, and WMAP
inner cut radius, as described in the lobe paper.  The photon index is
almost identical to that of model A, but the flux is somewhat lower
due to some of the core photons from model A being considered as being
emitted by the lobes in model B.  The results for model B can be seen
in Figure \ref{graySED}.

\subsection{Time Variability}

\begin{figure}
\epsscale{1.0}
\plotone{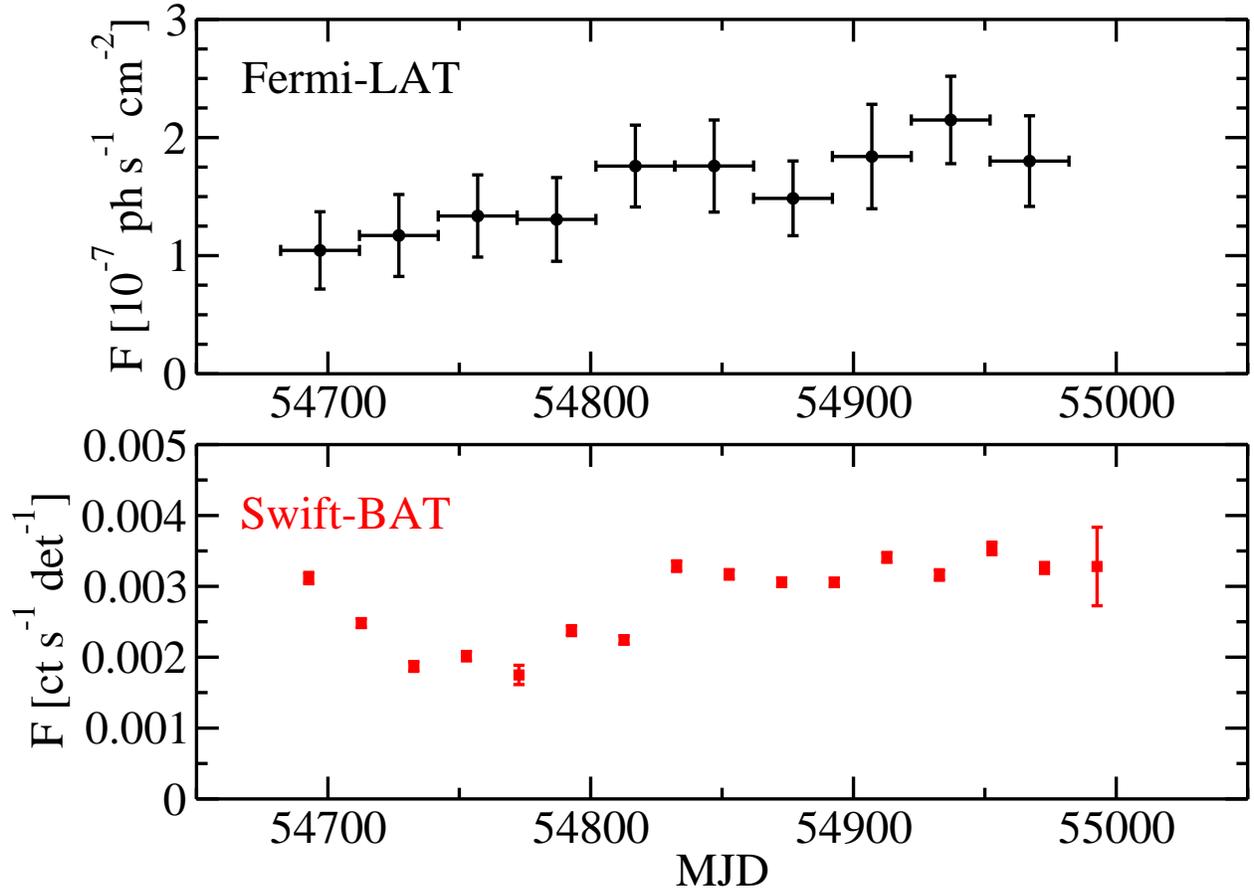}
\caption{(a) {\em Fermi}-LAT light curve of Cen~A without considering
lobe emission (Model A) in 30 day bins, with (b) simultaneous
lightcurve from {\em Swift}-BAT (14 day bins).  }
\label{figure-lightcurve}
\end{figure}

To quantify variability within the $\sim$10 month LAT observation, we
generated light curves in 30 and 15 day bins using the unbinned
likelihood analysis with {\tt gtlike}.  We performed the analysis
taking into account the lobe emission (i.e., Model B in \S\
\ref{LATspectralsection}). The power-law normalizations of the core
and background point sources are treated as free parameters, but the
photon indices of all sources and the normalizations of the lobes and
the diffuse background models are fixed to the values obtained in 200
MeV -- 30.0 GeV for the whole time region.
Figure~\ref{figure-lightcurve}a shows the light curve of the flux
(extrapolated down to $>100$ MeV) in 30 day bins.  The $\chi^2$
test results in $\chi^2$/d.o.f. = 0.98, and the light curve with 15
day bins gives $\chi^2$/d.o.f. = 0.89.  These are consistent with no
variability.  The time behavior of Cen~A is in contrast to large
variability of typical blazars in the MeV/GeV range, and similar to
that of Perseus A \citep{pera} and M87 \citep{m87}.

\section{ Other Contemporaneous Observations}
\label{obs}

Observations with several different instruments, both on the Earth and
in space, were made during the 10 months of LAT observations presented
here.  Cen~A was observed in the radio as part of the Tracking Active
Galactic Nuclei with Austral Milliarcsecond Interferometry (TANAMI)
program (Mueller et al.\ 2009; Ojha et al.\ 2009).  Data were taken
with two instruments on the {\em Swift} spacecraft \citep{gehrels04}
and two instruments on the {\em Suzaku} spacecraft
\citep{mitsuda07,koyama07,takahashi07}.  A summary of these
observations can be found in Table~\ref{tab-obs}, and descriptions are
given below.

\subsection{ Southern Hemisphere LBA Observations }

Cen~A was observed with VLBI on 2009 November 27/29, as part of the
TANAMI program using the five antennas of the Australian Long Baseline
Array (LBA), the 70\,m DSS-43 antenna at NASA's Deep Space Network at
Tidbinbilla, Australia, and two trans-oceanic telescopes TIGO (Chile)
and O'Higgins (Antarctica) of the International VLBI Service (IVS) for
Geodesy and Astrometry (the latter two participating at 8.4\,GHz,
only). The beam size achieved was (0.92 mas $\times$ 0.56 mas) at
8.4\,GHz and (1.68 mas $\times$ 1.25 mas) at 22.3\,GHz using natural
weighting.  These observations were part of the TANAMI monitoring of a
radio and $\gamma$-ray selected sample of 65 blazars at 8.4\,GHz and
22.3\,GHz with observations approximately every two months.

TANAMI data are correlated on the DiFX software correlator
\citep{deller07} at Curtin University in Perth, Western
Australia. Data inspection and fringe fitting was done with AIPS
(National Radio Astronomy Observatory's Astronomical Image Processing
System software).  The images were produced by applying the program
{\sc difmap} \citep{shepherd97}, using the {\sc CLEAN} algorithm. More
details about the data reduction can be found in \citet{ojha05}.

Data from the first epoch (November 2009) of TANAMI observations are
presented in \citet{ojha09}. Fig.\ \ref{figure-sed} includes the fluxs at
22.3\,GHz and 8.4\,GHz measured in 2009 November 27/29,
respectively. The total flux density, corresponding to the emission
distributed over the inner $\sim$ 120\,mas at 8.4\,GHz, is
$S_{\mathrm{total}}=3.90$\,Jy.  At 22.3\,GHz, a total VLBI flux
density of 3.2\,Jy is distributed over the inner $\sim 40$\,mas of the
jet, with very little emission on the counterjet side.

Via model fitting, we found a component with an inverted spectrum,
which is the brightest at both frequencies and which we identify with
the jet core. The core flux density is 0.92\,Jy at 8.4\,GHz and
1.54\,Jy at 22.3\,GHz. The core size is consistently modeled at both
frequencies to be (0.9--1.0)\,mas $\times$ (0.29--0.31)\,mas at the
same position angle of 53--55 degrees (see Ojha et al.\ 2009).

\subsection{ {\em Suzaku} Observations }

Cen~A was observed with {\em Suzaku} on 2009 July 20--21, Aug 5--6,
and Aug 14--16 with a total exposure of 150 ks, during which time the
flux approximately doubled.  We utilized data processed with version
2.4 of the pipeline {\em Suzaku} software, and performed the standard
data reduction: a pointing difference of $<1.5^{\circ}$, an elevation
angle of $>5^{\circ}$ from the earth rim, a geomagnetic cut-off
rigidity (COR) of $>$6 GV.  We did not use events from the time
the spacecraft entered the South Atlantic Anomaly (SAA) to $256$ s
after it left the SAA.  Further selection was applied: Earth
elevation angle of $>20^{\circ}$ for the X-ray Imaging Spectrometer
(XIS), COR$>$8 GV and the time elapsed from the SAA (T\_SAA\_HXD) of
$>$500 s for the Hard X-ray Detector (HXD).  The XIS response matrices
are created with {\tt xisrmfgen} and {\tt xissimarfgen}
\citep{ishisaki07}.  The HXD responses used here are {\tt
ae\_hxd\_pinhxnome5\_20070914.rsp} for the PIN and {\tt
ae\_hxd\_gsohxnom\_20060321.rsp} and {\tt
ae\_hxd\_gsohxnom\_20070424.arf} for the Gadolinium Silicate (GSO)
crystal.  The ``tuned'' ({\tt LCFIT}) HXD background files
\citep{fuka09} are utilized.  The detailed {\em Suzaku} analysis,
including GSO data and time variability, will be reported
elsewhere (Y.\ Fukazawa et al.\ 2010, in preparation).  The {\em
Suzaku} data were fit with a single absorbed power-law, which was
found to have a spectral index $\G=1.66\pm0.01$ with dust absorbing
column density $N_H = (1.08\pm0.01)\times10^{23}\ \cm^{-2}$.  The flux
in the 12 -- 76 keV band on 2009 July was $(1.23\pm0.01)\times10^{-9}\
\erg \s^{-1}\ \cm^{-2}\ \keV^{-1}$, about twice the flux measured by
{\em Suzaku} in 2005 \citep{mar07}.

\subsection{ {\em Swift}-XRT Observations }

\object{Cen~A} was observed on six days between 2009 Jan. 15 -- 28 for
a total exposure of 22 ksec (see Table \ref{tab-obs}).  The XRT
\citep{burrows05} data were processed with the XRTDAS software package
(v.~2.5.1) developed at the ASI Science Data Center (ASDC) and
distributed by the NASA High Energy Astrophysics Archive Research
Center (HEASARC) within the HEASoft package (v.~6.6).  Event files
were calibrated and cleaned with standard filtering criteria with the
{\em xrtpipeline} task using the latest calibration files available in
the \swift\ CALDB.

The XRT dataset was taken entirely in Windowed Timing mode.  For the
spectral analysis we selected events in the energy range 2--10 keV
with grades 0--2.  The source events were extracted within a box of
40x40 pixels ($\sim$94 arcsec), centered on the source position and
merged to obtain the average spectrum of Cen A during the XRT
campaign. The background was estimated by selecting events in a region
free of sources. Ancillary response files were generated with the {\em
xrtmkarf} task applying corrections for the PSF losses and CCD
defects.

The combined January X-ray spectrum is highly absorbed.  Hence it was
fitted with an absorbed power-law model with a photon spectral index
of $1.98\pm0.05$, an intrinsic absorption column of $(9.73\pm0.26)
\times 10^{22}\ \cm^{-2}$, in excess of the Galactic value of $8.1
\times 10^{20}\ \cm^{-2}$ in that direction \citep{kalberla05}.  The
average absorbed flux over the $2-10\ \keV$ energy range is
$(4.94\pm0.05) \times 10^{-10}\ \erg\ \cm^{-2}\ \s^{-1}$, which
corresponds to an unabsorbed flux of $9.15 \times 10^{-10}\ \erg\
\cm^{-2}\ \s^{-1}$.

The XRT spectrum included in the broadband SED was binned to ensure a
minimum of 2500 counts per bin and was de-absorbed by forcing the
absorption column density to zero in XSPEC, and applying a correction
factor to the original spectrum equal to the ratio of the de-absorbed
spectral model over the absorbed model.

\subsection{ {\em Swift}-BAT Observations } 

We used data from the Burst Alert Telescope (BAT) on board the Swift
mission to derive a 14--195\,keV spectrum of Cen-A contemporary to the
LAT observations.  The spectrum has been extracted following the
recipes presented in \citet[][]{ajello08,ajello09b}. This spectrum is
constructed by calculating weighted averages of the source spectra
extracted over short exposures (e.g.\ 300\,s).  These spectra are
accurate to the mCrab level and the reader is referred to
\cite{ajello09a} for more details.

\section{SED and Modeling}
\label{model-section}

\subsection{Spectral Energy Distribution}
\label{SEDsection}

The LAT spectrum of the core of Cen~A is shown in Fig.\
\ref{graySED}, extrapolated into the TeV regime, along with the HESS
spectrum observed between 2004 and 2008 \citep{aha09}.  Also shown is
the HESS spectrum scaled down by its source flux normalization uncertainty.
It seems that the LAT spectrum, with its statistical and systematic
errors, extrapolated to higher energies, is just barely consistent
with the HESS spectrum.  However, one should keep in mind that the
HESS and LAT spectra presented in this figure are not simultaneous,
although the HESS data did not show any signs of variability.
Additionally, $\g\g$ absorption makes it unlikely that the HESS and
LAT emission originate from the same region, which is explored below
(\S\ \ref{SSCmodel}).

Since the cores of many blazars have been shown to be $\g$-ray loud it
is plausible to assume that the radio core is the source of the
central $\g$-rays from Cen~A.  However, one should keep in mind that
the error circles of the {\em Fermi} and HESS \citep{aha09}
observations are consistent with emission from the inner lobes, jet
and radio core, so that these other regions could be sources of
$\g$-rays as well.  We construct the SED for the resolved sub-arcsec
and arcsec-scale core as compiled in \citet{mei07}, including their
mm/IR/optical observations from 2003--2005.  They have compiled
additional points from the 1990s and have applied an extinction
correction of $A_{\rm V} = 9$ mag to the optical and IR data.  We plot
historical data in the X-ray \citep{eva04}, hard X-rays
\citep{kinzer95,rot06,mar07}, COMPTEL \citep{ste98}, and the HESS TeV
$\g$-rays \citep{aha09}.  The {\em Swift} XRT and BAT, as well as {\em
Suzaku} data, corrected for Galactic dust as well as dust in NGC~5128,
discussed in \S\ \ref{obs}, were collected during time intervals
which overlap with much of the {\em Fermi}-LAT data.  Furthermore, we
add the simultaneous radio data of the TANAMI VLBI jet components.
All these are shown in Fig.\ \ref{figure-sed}.  The LAT data
points in Fig.\ \ref{figure-sed} are from Model B and include
statistical errors only.

\subsection{Synchrotron/Synchrotron self-Compton Model}
\label{SSCmodel}

Single-zone synchrotron/synchrotron self-Compton (SSC) models have
been very successful in explaining the multiwavelength (including
$\g$-ray) emission from BL Lac objects 
\citep[e.g.,][]{bloom96,tavecchio98}.  If FRIs are the
misaligned counterpart to BL Lacs, one would expect this model to
apply to them as well.  In this scenario the low energy, radio through
optical emission originates from nonthermal synchrotron radiation from
a relativistically moving spherical homogeneous plasma blob, and the
X-ray through VHE $\g$-rays from the Compton scattering of that
synchrotron radiation by electrons in the same blob.  The one-zone SSC
model has successfully fit the emission from the other {\em Fermi}-LAT
detected FRIs, \object{Perseus~A} \citep[NGC~1275;][]{pera} and
\object{M87} \citep{m87}, and has been successfully applied to
previous observations of \object{Cen~A} \citep{chi01}.  Here we apply
the single-zone SSC model to fit the recent multiwavelength
observations of \object{Cen~A}, particularly the {\em Fermi}-LAT and
HESS emission.

One can show (see Appendix \ref{analytic_appendix}) that, on the
assumption that all of the emission in the multiwavelength SED of the
Cen A core originates from the same region in a single zone SSC model,
$\g\g$ absorption gives the constraint on the Doppler factor
\begin{eqnarray}
\label{cenAconstrain1}
\delta_D \ge 5.3\ ,
\end{eqnarray}
where the Doppler factor is $\delta_D=[\G_j(1-\beta_j\mu)]^{-1}$, the
bulk Lorentz factor of the jet is $\G_j=(1-\beta_j^2)^{-1/2}$, $\beta_j
c$ is the speed of the jet, and $\theta=\cos^{-1}\mu$ is the angle of
the jet with respect to our line of sight.  Solving for $\G_j$ in
terms of $\delta_D$,
\begin{eqnarray}
\label{gammaquad}
\G_j = \frac{1 \pm \sqrt{1 - (1-\mu^2)(1+\delta_D^2\mu^2)} }
	{ \delta_D(1-\mu^2) }\ .
\end{eqnarray}
In order for $\G_j$ to be real, the quantity under the radical 
must be positive, which implies 
\begin{eqnarray}
\label{dDangle}
\delta_D \le \frac{1}{\sqrt{1-\mu^2}}\ = \csc\theta
\end{eqnarray}
\citep[e.g.,][]{urry95}.  For \object{Cen~A}, estimates of $\theta$
vary from $15^\circ$ to $80^\circ$ (see section \ref{cena-section}).
For the least constraining value, $\theta=15^\circ$,
\begin{equation}
\label{cenAconstrain2}
\delta_D \le 3.8\ .
\end{equation}

Clearly, the constraints (\ref{cenAconstrain1}) and
(\ref{cenAconstrain2}) are not compatible.  Thus, if the radio through
{\em Fermi} $\gamma$-ray data presented in Fig.\ \ref{figure-sed} are
synchrotron and SSC emission originating from the same region of the
jet, then {\em the HESS emission cannot originate from the same part
of the jet.  } Note also that the HESS emission cannot originate from
the same region of the jet, yet be emitted from a different mechanism
than SSC (say, Compton scattered accretion disk or dust torus
radiation) because even this radiation would be subject to the same
$\gamma\gamma$ attenuation by synchrotron photons.

If the VLBI jet core is assumed to be the origin of the high-energy
emission, the TANAMI core-size measurement can be used to calculate an
upper limit on the size of the $\g$-ray emitting region of $<0.017$\
pc = $5.3\times10^{16}$\ cm (\S\ 3.1).  This is consistent with the
VLBI observations of \citet{kellerman97} and \citet{horiuchi06},
and with a variability timescale of $t_v\sim 1$ day, given that
the emitting region radius $R_b$ is constrained by the variability
time by $R_b = \delta_D\ c\ t_v$.  This variability timescale is
consistent with the {\em Suzaku} observations, although it is not
clear that the {\em Suzaku} X-rays come from the same region as the
$\g$-rays.  Using this variability timescale and eqns
(\ref{SEDconstrain}) and (\ref{Bconstrain}), one gets $\delta_D=0.6$
and $B=6$ G.  More precise modeling \citep{finke08} gives the green
curve in Fig.\ \ref{figure-sed} with the model parameters in Table
\ref{modelparams}.  This curve demonstrates the emission can be fit
with a Doppler factor of unity.  This is consistent with a Lorentz
factor of unity or 7, a degeneracy which can be seen in eqn.\
(\ref{gammaquad}).  A stationary, nonrelativistic jet can explain the
entire SED, except the VHE emission.  This fit is similar to the
synchrotron/SSC fit by \citet{mei07} who fit similar data.  We further
note that a small change in $\delta_D$ leads to a large change in the
Lorentz factor.  This, combined with the uncertainty in the
inclination angle, leads to the fact that the Lorentz factor is not
well-constrained by modeling.  We also note that VLBI
observations show {\em apparent} motion with $\beta_{j,app}\sim 0.1$
\citep{tingay98}, implying $\G_j\ga 1.005$, which is also not a
particularly strong constraint.  

What if the hard X-ray emission originates from thermal Comptonization
near the disk, and not from jet emission?  If we assume the rest of
the high-energy SED is from the jet, then $\epsilon_{pk}^{SSC}=1$ and
$f_{pk}^{SSC}=9\times10^{-11}$ erg s$^{-1}$ cm$^{-2}$, so that eqns
(\ref{SEDconstrain}) and (\ref{Bconstrain}) give $\delta_D=2.4$ and
$B=0.6$ G for a variability timescale of 1 day.  More detailed
modeling gives the violet curve seen in Fig.\ \ref{figure-sed} 
with the parameters in Table \ref{modelparams}.  The larger Doppler
factor needed for this model requires a smaller angle to the line of
sight.  The Lorentz factor is again not strongly constrained, and
could plausibly be as high as $\G_j\sim8$ and still provide a good fit,
although this would push the parameters to their extremes.  This
model still under-predicts the HESS data.

Jet powers for these models are given in Table \ref{modelparams}.
The proton and pair content of the jet are not well known, so the
total jet power presented in Table \ref{modelparams} is for a pure
pair jet, and can be considerd a lower limit.  Even with 10--100 times
more energy in ions than leptons, the absolute jet power is far below
the Eddington luminosity for a $10^8 M_\odot$ black hole
($L_{Edd}=1.3\times10^{46}\ \erg\ \s^{-1}$).  For the green curve, the
parameters assume $\G=7$.  The jet power needed to inflate the giant
lobes of \object{Cen~A} in their lifetime, as inferred from the radio
spectral break, is $10^{43}$ erg s$^{-1}$ \citep{hardcastle09}.
This value is approximately consistent with the the green curve
model presented in Fig.\ \ref{figure-sed}.  

A possible explanation for the HESS observations is that the TeV
emission is produced by another blob.  We show in Fig.\
\ref{figure-sed} (brown curve) that another synchrotron/SSC-emitting
blob can produce the HESS emission without over-producing any of the
other multiwavelength data.  The parameters for this blob are in
Table \ref{modelparams}, although this fit is not unique and many
parameter sets would fit the HESS data and not contribute at other
wavelengths.  Other possible origins for the VHE emission are
discussed in \S\ \ref{VHEorigin}.

\subsection{Decelerating Jet Model}

Unification models for blazars suggest that FRII galaxies are FSRQs
with the jet viewed away from our line of sight, and similarly FRIs
are the parent population of BL Lacs.  In this case, one would expect
non-thermal emission from the cores of radio galaxies, de-beamed
compared to blazars.  However, the cores of FRIs seem brighter than
what is expected from simply de-beamed emission from BL Lacs, 
which implies the radio galaxy core emission is from a slower region
than that of BL Lacs, since the beaming angle is related to the bulk
Lorentz factor by $\theta_b\sim 1/\G_j$.  There are (at least) two
possible explanations for this: (1) the jet consists of a faster
``spine'', which is responsible for the on-axis blazar emission,
inside a slower outer ``sheath'', which would be responsible for the
off-axis emission seen in the cores of radio galaxies
\citep[e.g.,][]{chiaberge00}; and (2) a decelerating jet model where
the on-axis blazar emission is produced by a faster flow closer to the
black hole, and the off-axis $\g$-rays seen in radio galaxies are
produced by the slower flow farther out along the jet
\citep{georgan03b}.  

As an example, we provide a fit to the Cen~A SED using this
decelerating flow, as the blue curve in Fig.\ \ref{figure-sed}.  In
this model, the high energy emission is due to upstream Compton
scattering of synchrotron photons produced in the slower part of the
flow being scattered by energetic electrons in the faster, upstream
part of the flow.  The jet starts with a bulk Lorentz factor
$\Gamma_{j,max}=5$ and decelerates down to $\Gamma_{j,min}=2$ in a
length of $l=3 \times 10^{16}$ cm.  The injected power law electron
distribution, $n(\gamma)\propto \gamma^{-p}$ has an index $p=3.5$, and
extends from $\gamma_{min}=1600$, to $\gamma_{max}= 10^7$, and the
magnetic field at the inlet is $B=0.3 $ G.  Jet powers for this model
are similar to the one-zone SSC model fits presented in \S\
\ref{SSCmodel}, although this decelerating model fit is particle-
rather than magnetic field-dominated.  We also note that the
parameters used in this fit are not unique.

\section{Discussion}
\label{discussion}

\subsection{Origin of VHE $\g$-ray emission}
\label{VHEorigin}

Since the single blob model does not seem to be able to reproduce the
broadband SED of \object{Cen~A}, could something else be the origin of
the VHE $\gamma$-rays?  We have already shown that another blob
emitting synchrotron and SSC radiation could explain the HESS emission
without over-producing any of the other data (Fig.\ \ref{figure-sed}
brown curve).  \citet{lenain08} have presented a model with multiple
blobs, moving at different angles to the line of sight from a large
opening angle, to M87 and Cen~A (among other objects).  This model
does seem to be able to explain this SED \citep{lenain09}.  It has
also been suggested that absorbed $\gamma$-rays which create $e^+e^-$
pairs, creating an isotropic halo of electrons in the ISM which
Compton-scatter the host galaxy's starlight, leading to
isotropically-produced $\g$-rays \citep{sta03,sta06}.  The HESS data
do seem to match the \citet{sta06} predictions Cen~A with a galactic
magnetic field of 10 $\mu$G.  Compton-scattering off of leptons
accelerated by the supermassive black hole magnetosphere, similar to
particle acceleration in pulsars, has been proposed to explain the VHE
$\gamma$-ray radiation from M87 \citep{neronov07}.  This could also
explain the HESS data from \object{Cen~A} separate from the other
multiwavelength emission.  As we have noted earlier, what we designate
in this paper as the $\g$-ray ``core'' actually encompasses the radio
core, jet, and inner lobes of Cen~A.  This is also true for the HESS
emission.  \citet{croston09} have noted that a shock front observed in
X-rays in the southwest inner lobe could be a source of TeV $\g$-rays,
which seems consistent with these observations.

Finally, we note that the SED presented here is constructed from
non-simultaneous data.  Although {\em Fermi} and HESS $\gamma$-rays do
not show appreciable variability, they could still be variable on
longer timescales.  Perhaps for a good, simultaneous multiwavelength
SED, a one-zone synchrotron/SSC model could provide a good fit to all
of the data.  Probably the best way to discriminate between the above
models---simple SSC, Compton-scattering emission from a pair halo,
multiple blobs, etc.---is correlated variability between LAT
$\gamma$-rays and other bandpasses.  This emphasizes the importance of
simultaneous multiwavelength data.

\subsection{ Origin of UHE Cosmic Rays}

The Auger Observatory results indicate some UHECRs could be
originating from \object{Cen~A} (see \S\ \ref{cena-section}).  The
UHECRs could interact with photons at the source and in the
extragalactic background light leading to an observable signature in
the HESS band.  If the VHE $\gamma$-rays originate from cosmic rays
this could account for the discrepancy between HESS and {\em Fermi}
$\gamma$-rays.  Based on the green curve fit presented in Fig.\
\ref{figure-sed} we can analyze whether it is plausible for cosmic
rays to originate from \object{Cen~A}, keeping in mind that the
parameters of that model are not well constrained (\S\ 3).

The maximum energy to which cosmic rays can be accelerated is limited 
by the size scale of the emitting region and the highest energy 
they can reach before they are cooled.  The former constraint 
implies that the highest energy a cosmic ray can reach is 
\begin{equation}
\label{Esat2}
E_Z = 4\times10^{19}\ \frac{Z}{\phi}\ \left(\frac{B}{6.2\ G}\right)\ 
\left(\frac{t_{v}}{10^5\ \s}\right)\ \delta_D\ 
\left(\frac{\Gamma_j}{7.0}\right)\ \eV\ ,
\end{equation}
and the latter implies
\begin{equation}
\label{Esat1}
E_Z = 5.7\times10^{20}\ \sqrt{\frac{Z}{\phi}}\ \left( \frac{A}{Z} \right)^2\ 
	\left(\frac{B}{6.2\ G}\right)^{-1/2}\ \left(\frac{\G_j}{7.0}\right)\ \eV\ 
\end{equation}
\citep[e.g.,][]{hillas84,dermer10}, where $\phi\approx1$ is the
acceleration efficiency factor, and $e$ is the elementary charge, $Z$
is the atomic number, and $A$ the atomic mass of the ion.  Note that
these timescales, and all quantities expressed above, are in the frame
comoving with the blob, although for the particular model considered
here, $\delta_D$=1 so this is not important.

We assume all parameters have values from the green curve model.
Thus, it seems for this model that it is unlikely that protons will be
accelerated to energies above $\approx 4\times10^{19}$ eV, although it
is possible for heavier ions to be accelerated this high before they
are disintegrated by interacting with infrared photons from the
\object{Cen~A} core.  The threshold energy for photomeson interaction
with peak synchrotron photons is similar to $E_Z$.  This process could
create observational signatures from secondary emission
\citep[e.g.,][]{kachel09}, as well as convert protons to neutrons,
which can escape as cosmic rays \citep{dermer09}.  Again, we note that
this result is strongly model-dependent, and the parameters of this
model are not strongly constrained, so this limit should not be taken
too seriously.  For example, a small change in the Doppler factor
would have little effect on the model fit, but would require a large
change in the bulk Lorentz factor, $\G_j$.  A large change in $\G_j$
would significantly affect the highest energy to which particles could
be accelerated, as seen in eqns (\ref{Esat1}) and (\ref{Esat2}).
Furthermore, if we are viewing a slower sheath, UHE cosmic rays could
be accelerated in the faster spine beamed away from our line of sight,
which could have significantly different parameters.  Acceleration of
protons up to $10^{20}\ \eV$ requires jet powers of $P_j\ga10^{46}\
\erg\ \s^{-1}$, which may take place in occasional flaring activities
in \object{Cen~A} \citep{dermer09}.

\section{Summary}
\label{summary}

We have reported on observations of \object{Cen~A} with the LAT
instrument on board the {\em Fermi} Gamma-Ray Space Telescope.  This
instrument's excellent angular resolution compared to other $\g$-ray
detectors at MeV--GeV energies makes it possible for the first time to
separate the lobe and core emission.  The LAT observations have been
supplemented with simultaneous observations from {\em Suzaku}, {\em
Swift}, the Australia Telescope Long Baseline Array, and a variety of
non-simultaneous data, including those from HESS.  Our results are as
follows:

\begin{enumerate}
\item
The LAT-detected core position is consistent with Cen~A's VLBI
core \citep{ma98} and previous EGRET observations \citep{har99}.
\item
With 10 months of LAT exposure, we find the core flux $>100$
 MeV to be (1.50$\pm0.25_{stat}\pm0.37_{sys}$) \gunit\  and the
 spectral index in the $0.2-30$ GeV range to be
 $\Gamma$=$2.67\pm0.10_{stat}\pm0.08_{sys}$, consistent with the
 EGRET \citep{har99} and the previously-reported 3-month LAT
 detection \citep{lbas}.
\item
Extrapolated to higher energies, the LAT spectrum is barely 
consistent with the HESS spectrum \citep{aha09} only if the 
HESS spectrum is lowered in flux by its normalization error.
\item
A single zone SSC model can explain all of the multiwavelength
emission from the core except for the non-simultaneous HESS emission.
It is not possible to fit the entire SED, including the HESS emission,
with a single zone Compton-scattering model due to internal
$\gamma\gamma$ absorption effects.
\item
Modeling results are consistent with suggestions by Chiaberge et al.
that we are seeing $\g$-rays from a different origin that we would if
were were looking down the jet.  This could be explained by a spine in
sheath \citep{chiaberge00} or decelerating jet scenario
\citep{georgan03b}.
\end{enumerate}

\acknowledgments

The \textit{Fermi} LAT
Collaboration acknowledges generous ongoing support from a number of
agencies and institutes that have supported both the development and
the operation of the LAT as well as scientific data analysis.  These
include the National Aeronautics and Space Administration and the
Department of Energy in the United States, the Commissariat \`a
l'Energie Atomique and the Centre National de la Recherche
Scientifique / Institut National de Physique Nucl\'eaire et de
Physique des Particules in France, the Agenzia Spaziale Italiana and
the Istituto Nazionale di Fisica Nucleare in Italy, the Ministry of
Education, Culture, Sports, Science and Technology (MEXT), High Energy
Accelerator Research Organization (KEK) and Japan Aerospace
Exploration Agency (JAXA) in Japan, and the K.~A.~Wallenberg
Foundation, the Swedish Research Council and the Swedish National
Space Board in Sweden.

Additional support for science analysis during the operations phase is
gratefully acknowledged from the Istituto Nazionale di Astrofisica in
Italy and the Centre National d'\'Etudes Spatiales in France.

We are grateful to the anonymous referee for useful comments
which have improved the manuscript.  We thank N.~Odegard for providing
us with the WMAP image and the Suzaku team for their calibration and
satellite operation.  We also acknowledge the Swift Team and the
Swift/XRT monitoring program efforts, as well as Swift analysis
supported by NASA grants NNX08AV77G and NNX09AU07G.


\appendix

\section{$\g\g$ Absorption Constraint on the Doppler Factor of Cen~A}
\label{analytic_appendix}

\begin{figure}
\epsscale{0.8} 
\plotone{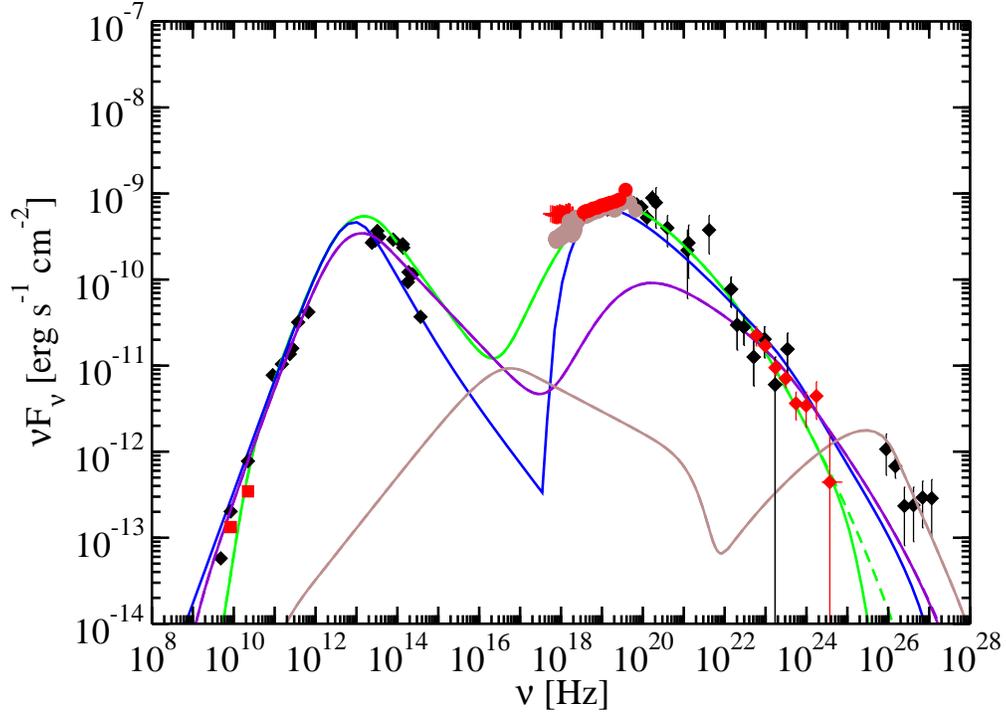} 
\caption{ The SED of the Cen~A core with model fits.  Colored symbols
are observations between August and May 2009, the epoch of the LAT
observations.  These include observations of, from low to high
frequency: the TANAMI VLBI (red squares), {\em Swift}-XRT (red
crosses), {\em Suzaku} (brown circles), {\em Swift}-BAT (red circles),
and {\em Fermi}-LAT (red diamonds).  Black symbols are archival data,
\citep{marconi00} including HESS observations \citep{aha09}.  Curves
are model fits to nuclear region of Cen~A.  The green curve is a
synchrotron/SSC fit to the entire data set.  The dashed green curve
shows this model without $\gamma\gamma$ attenuation.  The violet curve
is a similar fit but is designed to under fit the X-ray data, and the
brown curve is designed to fit the HESS data while not over-producing
the other data in the SED.  The blue curve is the decelerating jet
model fit \citep{georgan03b}.  See Table \ref{modelparams} for the
parameters of these model curves.  }
\label{figure-sed}
\end{figure}

In the SSC model, the Doppler factor, $\delta_D$, and comoving,
tangled, isotropic magnetic field strength, $B$, may be estimated from
the dimensionless peak energy and $\nu F_{\nu}$ flux, $\epsilon_{pk}$
and $f_{pk}^{syn}$ of the synchrotron and SSC components,
respectively, observed in the SED.  Assuming the comoving blob
size can be constrained by $R_b^\prime=t_v\delta_Dc/(1+z)$, this gives
\begin{eqnarray}
\label{SEDconstrain}
\delta_D = 1.6\ 
	\left(\frac{\epsilon_{pk}^{SSC}}{1}\right)^{1/2}
	\left(\frac{10^{-7}}{\epsilon_{pk}^{syn}}\right)
	\left(\frac{D}{10^{25}\ \cm}\right)^{1/2}
	\left(\frac{1\ \dday}{t_v}\right)^{1/2}
\\ \nonumber \times
	\left(\frac{f_{pk}^{syn}}{10^{-10}\ \erg\ \s^{-1}\ \cm^{-2}}\right)^{1/2}
	\left(\frac{10^{-10}\ \erg\ \s^{-1}\ \cm^{-2}}{f_{pk}^{SSC}}\right)^{1/4}
\end{eqnarray}
\begin{eqnarray}
\label{Bconstrain}
B = 0.26\ G\ 
	\left(\frac{t_v}{1\ \dday}\right)^{1/2}
	\left(\frac{10^{25}\ \cm}{D}\right)^{1/2}
	\left(\frac{\epsilon_{pk}^{syn}}{10^{-7}} \right)^3
	\left(\frac{1}{\epsilon_{pk}^{SSC}}\right)^{3/2}
\\ \nonumber \times
	\left(\frac{f_{pk}^{SSC}}{10^{-10}\ \erg\ \s^{-1}\ \cm^{-2}}\right)^{1/4}
	\left(\frac{10^{-10}\ \erg\ \s^{-1}\ \cm^{-2}}{f_{pk}^{syn}}\right)^{1/2}
\end{eqnarray}
\citep{ghisel96} where $t_v$ is the variability timescale and $D$ is
the distance to the source.  The Doppler factor,
$\delta_D=[\Gamma_j(1-\beta_j\mu)]^{-1}$ where the Bulk Lorentz factor is
$\Gamma_j=(1-\beta_j^2)^{-1/2}$, $\beta_j c$ is the speed of the jet, and
$\theta=\cos^{-1}\mu$ is the angle of the jet with respect to our line
of sight.  In order for $\gamma$-rays to escape an emission region,
the $\gamma\gamma \rightarrow e^+e^-$ absorption optical depth,
$\tau_{\gamma\gamma}$, cannot be too large.  Assuming the $\nu
F_{\nu}$ synchrotron flux, $f_\epsilon^{syn}$, is given by a broken
power law, then for $\tau_{\gamma\gamma}<1$ for a photon with
dimensionless energy $\epsilon_\gamma$, this implies
\begin{eqnarray}
\label{ggconstrain}
\delta_D \ge \Biggr[10^3\times 
	2^{A-1}\ (1+z)^{2-2A}
	\left(\frac{\epsilon_{\gamma}}{10^{7}}\right)
	\left(\frac{D}{10^{25}\ \cm}\right)^2
\\ \nonumber \times
	\left(\frac{f^{syn}_{\epsilon^{-1}_\gamma}}
	{10^{-10}\ \erg\ \s^{-1}\ \cm^{-2}}\right)
	\left(\frac{1\ \dday}{t_v}\right)\ 
	\Biggr]^{\frac{1}{6-2A}}
\end{eqnarray}
\citep{dondi95} where $f_\epsilon^{syn}\propto\epsilon^A$ and $A$ is
the index of the synchrotron spectrum below the break for
$$
\epsilon_\gamma^{-1}< \frac{(1+z)^2 \epsilon_{brk}}{2\delta_D}
$$
and above the break for
$$
\epsilon_\gamma^{-1}>\frac{(1+z)^2 \epsilon_{brk}}{2\delta_D}\ .
$$
Solving eqn. (\ref{SEDconstrain}) for $t_v$ and inserting this into
eqn. (\ref{ggconstrain}), one gets the constraint
\begin{eqnarray}
\label{constrain2}
\delta_D \ge 4.4\ \Biggr[\ 2^{1-A} (1+z)^{2-2A} 
	\left(\frac{\epsilon_{\gamma}}{10^{7}}\right)
	\left(\frac{D}{10^{25}\ \cm}\right)
	\left(\frac{f^{syn}_{\epsilon^{-1}_\gamma}}
	{10^{-10}\ \erg\ \s^{-1}\ \cm^{-2}}\right)
\\ \nonumber \times
	\left(\frac{\epsilon_{pk}^{syn}}{1\times10^{-7}}\right)^2
	\left( \frac{1}{\epsilon_{pk}^{SSC}} \right)
	\left(\frac{f_{pk}^{syn}}{10^{-10}\ \erg\ \s^{-1}\ \cm^{-2}}\right)^{1/2}
	\left(\frac{10^{-10}\ \erg\ \s^{-1}\ \cm^{-2}}{f_{pk}^{SSC}}\right)\ 
	\Biggr]^{1/4}\; .
\end{eqnarray}
For \object{Cen~A}, $D=3.7$ Mpc $=1.1\times10^{25}$ cm, and
$z\approx 0$.  The spectral parameters can be obtained from the SED of
the core of \object{Cen~A} (see Fig.\ \ref{figure-sed}):
$\epsilon_{pk}^{syn}=1.6\times10^{-7}$, $\epsilon_{pk}^{SSC}=0.3$,
$f_{pk}^{syn} = 3\times10^{-10}$ erg\ s$^{-1}$\ cm$^{-2}$, and
$f_{pk}^{SSC} = 9\times10^{-10}$ erg\ s$^{-1}$\ cm$^{-2}$.  Note that
here we assume that the X-ray data is from the jet; see above.  Below
the break in the synchrotron spectrum, $A\approx 0.5$, and above
$A\approx -1$.  The highest energy photon bin in the HESS spectrum is
$\epsilon_\gamma=8\times10^6$, so that
$f^{syn}_{\epsilon^{-1}_\gamma}=2\times10^{-10}$ erg\ s$^{-1}$\
cm$^{-2}$.  These values give the constraint
$$
\delta_D \ge 5.3\ ,
$$
which is equation (\ref{cenAconstrain1}).  

{}

\clearpage

\begin{deluxetable}{lccc}
\tabletypesize{\scriptsize}
\tablecaption{
Summary of multiwavelength observations.
}
\tablewidth{0pt}
\tablehead{
\colhead{Instrument} &
\colhead{Observation date} &
\colhead{Exposure time } &
\colhead{Frequency/Energy range } 
}
\startdata
Australian LBA and IVS & 2009 Nov.\ 27& 3.6 ks & 22.3 GHz \\
                       & 2009 Nov.\ 29 & 3.6 ks & 8.4 GHz \\
{\em Suzaku} XIS  & 2009 Jul.\ 20 -- Aug. 16 & 150 ks & 0.4 -- 10 keV \\
{\em Suzaku} HXD-PIN  & 2009 July 20 -- Aug. 16 & 150 ks & 10 -- 70 keV \\
{\em Swift} XRT   &  2009 Jan.\ 15 -- 28 & 22 ks & 0.2 -- 10 keV \\ 
{\em Swift} BAT   & 2008 Aug.\  -- 2009 May & 1.9 Ms & 14 -- 200 keV \\
{\em Fermi} LAT & 2008 Aug.\ 4  -- 2009 May 31 & 10 Months & 0.2--30 GeV \\
\enddata
\label{tab-obs}
\end{deluxetable}

\begin{deluxetable}{lccccc}
\rotate
\tabletypesize{\scriptsize}
\tablecaption{
Model Parameters.  
}
\tablewidth{0pt}
\tablehead{
\colhead{Parameter} &
\colhead{Symbol} &
\colhead{Green\tablenotemark{1} } &
\colhead{Blue\tablenotemark{2} } & 
\colhead{Violet\tablenotemark{3} } & 
\colhead{Brown\tablenotemark{4} }
}
\startdata
Bulk Lorentz Factor & $\Gamma_j$	& 7.0 & $5\rightarrow 2$ & 3.7 & 2.0\\
Doppler Factor & $\delta_D$       & 1.0 & $1.79\rightarrow 1.08$ & 3.9 & 3.1 \\
Jet Angle & $\theta$ & $30^\circ$ & $25^\circ$& $15^\circ$ & $15^\circ$ \\
Magnetic Field [G]& $B$ & 6.2  & 0.45 & 0.2 & 0.02 \\
Variability Timescale [sec]& $t_v$ & $1.0 \times 10^5$  & & $1\times10^5$  & $1\times10^5$ \\
Comoving blob size scale [cm]& $R_b$ & $3.0\times10^{15}$  & $3\times10^{15}$ & $1.1\times10^{16}$  & $9.2\times10^{15}$  \\
\hline
Low-Energy Electron Spectral Index & $p_1$       & 1.8 & 3.2 & 1.8 & 1.8  \\
High-Energy Electron Spectral Index  & $p_2$       & 4.3 & & 4.0 & 3.5  \\
Minimum Electron Lorentz Factor & $\gamma_{min}$  & $3\times10^2$ & $1.3\times10^3$ & $8\times10^2$ & $8\times10^2$ \\
Maximum Electron Lorentz Factor & $\gamma_{max}$  & $1\times10^8$ & $1\times10^7$ & $1\times10^8$ & $1\times10^8$ \\
Break Electron Lorentz Factor & $\gamma_{brk}$ & $8\times 10^2$ &  & $2\times10^3$ & $4\times10^5$ \\ 
\hline
Jet Power in Magnetic Field [$\erg\ \s^{-1}$] & $P_{j,B}$ & $6.5\times10^{43}$ & $1.7\times10^{41}$ & $2.7\times10^{41}$ & $4.3\times10^{38}$  \\
Jet Power in Electrons [$\erg\ \s^{-1}$] & $P_{j,e}$  & $3.1\times 10^{43}$ & $3.1\times10^{42}$ & $2.3\times10^{42}$ & $7.0\times10^{40}$ \\
\enddata
\tablenotetext{1}{SSC Model}
\tablenotetext{2}{Decelerating Jet Model \citep{georgan03b} }
\tablenotetext{3}{SSC Model excluding X-rays}
\tablenotetext{4}{SSC Fit to HESS data only}
\label{modelparams}
\end{deluxetable}

\clearpage

\end{document}